# On the Rate Distortion Function of Certain Sources with a Proportional Mean-Square Error Distortion Measure

Jacob Binia, *Member, IEEE*
Email: biniaja@netvision.net.il

*Abstract*— New bounds on the rate distortion function of certain non-Gaussian sources, with a proportional-weighted mean-square error (MSE) distortion measure, are given. The growth, g, of the rate distortion function, as a result of changing from a non-weighted MSE distortion measure to a proportional-weighted distortion criterion is analyzed. It is shown that for a small distortion, d, the growth, g, and the difference between the rate distortion functions of a Gaussian memoryless source and a source with memory, both with the same marginal statistics and MSE distortion measure, share the same lower bound. Several examples and applications are also given.

## I. INTRODUCTION

The rate distortion function for Gaussian processes, as well as for certain non-Gaussian processes, with a mean-square error (MSE) distortion measure is well known (e.g. [1]-[4]). In order to preserve signal's fine and more fragile features that might disappear under source coding that aims only at minimizing the mean-square error energy, Sakrison [5] suggested the following generalization on the MSE distortion measure to a class of weighted MSE distortion measures. Let $D_{A,I_T}$ denotes the weighted MSE distortion

$$D_{A,I_T} = E\{\frac{1}{T}\|A(X - \hat{X})\|_{I_T}^2\}, \qquad (1)$$

where $X, \hat{X}$ are the source signal and its reconstruction respectively, A is a linear, time-invariant operator that is bounded from below, $I_T$ is a T seconds interval and

$$\|f(t)\|_{I_T}^2 = \int_{I_T} f^2(t)dt .$$

As usual the distortion and the rate are calculated by taking the limits for $T \to \infty$. Let X be a stationary process with a power spectral density $S_X(f)$ and let A(f) be the transfer function of the operator A. Then, under some conditions, the following parametric expressions for the rate distortion were proven in [5]:

$$D_A(\mu) = \mu \int_{E(\mu)} df + \int_{E^c(\mu)} |A(f)|^2 S_X(f)df \qquad (2)$$

$$R_A(\mu) = -\frac{1}{2} \int_{E(\mu)} \ln[\mu / |A(f)|^2 S_X(f)]df . \qquad (3)$$

In (2), (3) $E(\mu)$ is the set of frequencies for which $|A(f)|^2 S_X(f) \geq \mu$ and $E^c(\mu)$ is its complement.

The meaning of the result above is that the rate distortion function of the stationary time-continuous Gaussian source X with a weighted MSE distortion measure is equal to the rate distortion function with a regular (non-weighted) MSE distortion measure of a stationary Gaussian source Y with power spectral density $S_Y(f) = |A(f)|^2 S_X(f)$. The frequency-weighted MSE criterion for time-continuous Gaussian sources is also employed in [6]-[8] with similar results. However, no results for a specific example of weighting transformation were given.

In this paper we focus on a special form of weighted MSE distortion measure. A will be defined so that the condition $|A(f)|^2 S_X(f)$ equals a constant is fulfilled for each frequency f. This criterion will cause to the power spectral density of the (non-weighted) *error* signal, as a function of f, to be proportional to that of the signal source X, for each f. Our new results will also apply to certain non-Gaussian sources.

In section II the rate distortion functions for stationary time-discrete, and for bandlimited time-continuous processes are analyzed. Bounds on the rate distortion function under the proportional-weighted MSE distortion measure, as well as asymptotic behavior for a small distortion d, are given. The case of unlimited-bandwidth, time-continuous sources is also treated. Proofs of main results are given in the appendix.

## II. R(d) FOR STATIONARY SOURCES

### A. R(d) for time-discrete stationary sources

Let $X = \{X_t, t = 0, \pm 1, \dots\}$ be a zero mean stationary sequence with $E[X^2(t)] = S < \infty$ and with a power spectral density function $\Phi_X(f), 1/2 \leq f \leq 1/2$. Let $D(X\|\tilde{X}) < \infty$ be the divergence rate

$$D(X\|\tilde{X}) = \lim_{n\to\infty}\frac{1}{n}D((X_1,...,X_n)\|(\tilde{X}_1,...,\tilde{X}_n))$$

where $\tilde{X},(\tilde{X}_1,...,\tilde{X}_n)$ are Gaussian with the same covariance as that of $X,(X_1,...,X_n)$ respectively. Define the linear time-invariant operator A by

$$|A(f)|^2\,\Phi_X(f) = S. \tag{4}$$

First, we use the above classical definition of weighted-distortion measure using the linear transformation (4) and consider only spectral densities $\Phi_X(f)$ such that $A^{-1}$ exist. This restriction will be removed in the end part of this paragraph. We denote the distortion $d_n$ between any source-reconstruction words pair by

$$d_n = E\{\frac{1}{n}\left\|A(X-\hat{X})\right\|_n^2\}\ . \tag{5}$$

In (5) $\|\bullet\|_n^2$ symbolizes the usual norm in the Euclidean n space and $d = \lim_{n\to\infty} d_n$. Then, the rate distortion function $R(d)$ of X satisfies the following theorem

*Theorem 1:* For $0 < d \le S$

$$\frac{1}{2}\ln\frac{S}{d} - D(X\|\tilde{X}) \le R(d) \le \frac{1}{2}\ln\frac{S}{d}\ . \tag{6}$$

If $D(X\|\tilde{X}) < \infty$, then as $d \to 0$

$$R(d) = \frac{1}{2}\ln\frac{S}{d} - D(X\|\tilde{X}) + o(1)\ , \tag{7}$$

where o(1) denotes a function that goes to zero as d goes to zero.

If $A^{-1}$ exists then the proof of the theorem is straightforward. Each pair $(X,\hat{X})$ can be transformed to another pair $(AX,\widehat{AX})$ and reconstructed by operating the inverse without loss of information. According to the definition (4) of A the power density function of the process Y=AX is "white". Therefore the proportional-distortion rate distortion function of X is given by the non-weighted MSE rate distortion function of the memoryless process Y with same d. In this case the rate distortion function of Y under the (non-weighted) MSE distortion measure is received by a constant distortion d for each frequency f, $-\frac{1}{2}\le f\le\frac{1}{2}$ (due to error water filling law). Therefore, the weighted error spectrum of d (4), (5) with $n\to\infty$, as a function of f, is proportional to $\Phi_X(f)$ (see figure 1).

*Example 1:* First-order Markov source

We consider a first-order Gaussian Markov source X which is characterized by an exponentially decaying memory (see [1, example 4.5.2.2]). For convenience we assume S=1. The power spectral density of X is given by:

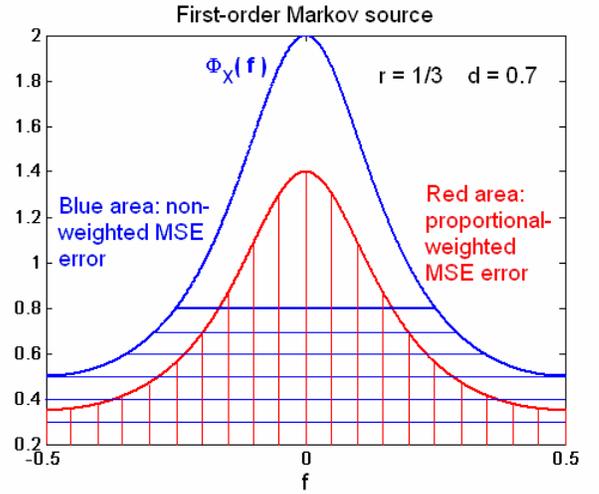

Fig. 1: Errors power spectral densities of first-order Markov source

$$\Phi_X(f) = \frac{1-r^2}{1-2r\cos(2\pi f)+r^2}\ ;\quad -\frac{1}{2}\le f\le\frac{1}{2} \tag{8}$$

where $-1 < r < 1$. A graph of $\Phi_X(f)$ for r=1/3 is shown in Figure1. For d=0.7 the blue area denotes the error spectrum due to the non-weighted MSE distortion measure. The red area denotes the error spectrum due to the proportional-weighted MSE distortion measure $(=0.7*\Phi_X(f)\,)$.

The rate distortion function for the non-weighted MSE distortion measure for $d\le (1-r)/(1+r)$, is given by [1, 4.5.32]

$$R(d) = \frac{1}{2}\ln\frac{1-r^2}{d}\ . \tag{9}$$

The "cost" in terms of source encoding rate required in order to reconstruct the data using a *proportional* weighted distortion MSE measure can now be seen. The following expression gives the growth, g, from the rate distortion (6) to the rate distortion (9), both with same d, for the Gaussian case $(D(X\|\tilde{X}) = 0\,)$ and S=1,

$$g = -\frac{1}{2}\ln(1-r^2)\ . \tag{10}$$

As it can be seen, depending on the memory decaying factor r, the growth of the rate varies between zero (memoryless source) and infinity.

One can come to a similar conclusion in the case of non-Gaussian sources, at least for small d. The generalized Shannon lower bound for the non-weighted MSE [1, 4.6.14], when expressed in terms of divergence, is given by

$$R(d) \ge \frac{1}{2}\ln\frac{1}{d} - D(X\|\tilde{X}) + \frac{1}{2}\int_{-1/2}^{1/2}\ln\Phi_X(f)df\ .$$

Therefore, for small d (7) yields

$$g \ge \frac{1}{2}\ln S - \frac{1}{2}\int_{-1/2}^{1/2}\ln\Phi_X(f)df + o(1)\ . \tag{11}$$

For a memoryless, non-Gaussian source the lower bound (11) behaves like o(1) for small d, but becomes higher for sources with memory.

In [9] it was shown that for any time-discrete stationary Gaussian source with memory, the rate distortion R(d) satisfies

$$R^*(d) + \frac{1}{2}\ln S - \int_{-1/2}^{1/2}\ln\Phi_X(f)df \le R(d) \le R^*(d)$$

where $R^*(d)$ is the rate of the memoryless source $X^*$ with the same marginal statistics as X. Hence, the growth of the rate distortion function when passing from a non-weighted MSE criterion to a proportional-weighted criterion, and the growth of the rate distortion function when passing from a Gaussian memoryless source to a source with memory and with the same marginal statistics, share the same lower bound (for small d).

*A generalization*

Let us suggest another approach to the solution of the above problem concerning the proportional- weighted MSE distortion measure. Suppose we dictate in advance an error spectral density function that is proportional to the source power density, for each frequency f. Let $\lambda_k^{(n)}$ $k=1,...n$ be the eigenvalues of the correlation matrix $\phi_n(X)$ of any n-sequence belongs to X. Let $\gamma_k^{(n)}$ $k=1,...n$ be the eigenvalues of the correlation matrix $\phi_n(e)$ of any n-sequence belongs to the *error* process $e=(X-\hat{X})$.

Then, using a result by Grenander and Szego (e.g. see [1] p. 116), the eigenvalues $\lambda_k^{(n)}, \gamma_k^{(n)}$ $k=1,...n$ can be expressed for some $f_k^{(n)}$ $k=1,...,n$ by $\Phi_X(f_k^{(n)})$, $\Phi_e(f_k^{(n)})$ respectively (same $f_k^{(n)}$ since the spectral densities are proportional). Also, $\Phi_e(f)=\frac{d}{S}\Phi_X(f)$ implies

$$\gamma_k^{(n)}=\frac{d}{S}\lambda_k^{(n)} \quad k=1,...n\,.$$

By transferring coordinates to the principal axes $(X_1,...,X_n) \to (X_1^*,...,X_n^*)$ we get

$$\lim_{n\to\infty}\frac{1}{n}\sum_{k=1}^{n}E[(X_k^*-\hat{X}_k^*)^2]=\lim_{n\to\infty}\frac{1}{n}\sum_{k=1}^{n}\gamma_k^{(n)}= \\ \lim_{n\to\infty}\frac{1}{n}\sum_{k=1}^{n}\frac{d}{S}\lambda_k^{(n)}=\lim_{n\to\infty}\frac{1}{n}\frac{d}{S}\sum_{k=1}^{n}\lambda_k^{(n)}=d. \quad (12)$$

Therefore, the rate distortion function can be calculated by finding the minimum information rate between $(X_1^*,...,X_n^*)$ and $(\hat{X}_1^*,...,\hat{X}_n^*)$, where $E[X_k^*]^2=\lambda_k^{(n)}$, $k=1,...,n$, subject to the condition $E[(X_k^*-\hat{X}_k^*)^2]=\gamma_k^{(n)}$, $k=1,...,n$.

The following expression for the rate distortion function in the Gaussian case is straightforward:

$$R(d)=\lim_{n\to\infty}\frac{1}{2n}\sum_{k=1}^{n}\ln\frac{S}{d}=\frac{1}{2}\ln\frac{S}{d}\,.$$

The bounds (6), as well as the asymptotic behavior for small d (7), for the non-Gaussian case are proved in the appendix. Hence, we do not need to pass through the transformation A in order to define the proportional MSE distortion measure or to prove theorem 1. Moreover, no existence of $A^{-1}$ is needed (e.g. the power density function of the source X can possesses zero values).

*B. R(d) for time-continuous bandlimited sources*

Let $x=\{x_t, -\infty<t<\infty\}$ be a zero mean stationary process with $E[x^2(t)]=S<\infty$ and power density function $S_x(f)$ such that $S_x(f)=0$ $|f|>B$. Lower case x is used to denote time-continuous process and upper case X to denote the sampled, time-discrete sequence $X=\{x(kT_s), k=0,\pm1,...\}$, where $1/T_s=2B$ is the sampling rate.

Let the linear time-invariant operator A be defined by

$$|A(f)|^2 S_x(f)=S/2B\,. \quad (13)$$

Again, we can consider the proportional-weighted MSE distortion measure in two forms. Using the linear operator A, we denote the distortion $d_T$ between any T- seconds source-reconstruction processes pair by

$$d_T=E\{\frac{1}{T}\|A(x-\hat{x})\|_T^2\}, \quad (14)$$

where $\|\bullet\|_T^2$ symbolizes the usual norm in $L^2[0,T]$ and by d, $d=\lim_{T\to\infty}d_T$, the distortion rate. More generally and without the need to restrict A, we demand that the power density of the *error* process $e=(x-\hat{x})$ , $S_e(f)$, satisfies for each frequency f

$$S_e(f)=\frac{d}{S}S_x(f)\,. \quad (15)$$

The following theorem is a straightforward consequence of theorem 1. Its proof is given in the appendix.

*Theorem* 2: For $0<d\le S$

$$B\ln\frac{S}{d}-2BD(X\|\tilde{X})\le R(d)\le B\ln\frac{S}{d}\,. \quad (16)$$

If $D(X\|\tilde{X})<\infty$ , then, as $d\to0$,

$$R(d)=B\ln\frac{S}{d}-2BD(X\|\tilde{X})+o(1)\,, \quad (17)$$

where o(1) denotes a function that goes to zero as d goes to zero.

*Corollary:* Suppose that a source x with $E[x^2(t)]=S<\infty$ and with power density $S_x(f)$ such that $S_x(f)=0$ $|f|>B$ is passed to a destination through a white, band-limited Gaussian channel that possesses signal to noise ratio SNR, and

shares the same bandwidth B as that of x. Then, even for the best source and channel encoders, and the best channel and source decoders

$$\frac{d}{S} \geq \frac{1}{(1+\text{SNR}) \exp[2 D(X \| \tilde{X})]} \ . \qquad (18)$$

In particular, if the source is Gaussian

$$\frac{d}{S} \geq \frac{1}{(1+\text{SNR})}. \qquad (19)$$

A numerical example: The minimum SNR that is required for the above Gaussian link in order to maintain a proportional distortion of 10% is 9 (9.5 dB).

Let us check the growth g of the rates when passing from a non-weighted MSE distortion to a proportional-weighted MSE distortion. Bounds on the rate distortion function of non-weighted, time-continuous stationary bandlimited source were given in [1, (4.6.46)]. The lower bound when expressed in terms of divergence rate and the power density function $\Phi_X(f)$ of the sampled process X, is in the following form:

$$R(d) \geq B \ln \frac{S}{d} - 2B\, D(X \| \tilde{X}) +$$
$$B \int_{-1/2}^{1/2} \ln \Phi_X(f) df - B \ln S. \qquad (20)$$

Then, from (17) and (20), for small d

$$g \geq B \ln S - B \int_{-1/2}^{1/2} \ln \Phi_X(f) df + o(1) \ . \qquad (21)$$

Again, comparing (21) with the difference between the rate distortion function of a discrete-time Gaussian source and its memoryless equivalent with the same marginal statistics [9], one can come to the following conclusion:

The growth g for small d is at least the difference between the rates of the sampled discrete-time Gaussian source with power density function $\Phi_X(f)$ and its equivalent memoryless, multiplied by minus B.

### C. R(d) for time-continuous, infinite- bandwidth sources

Let $x = \{x_t, -\infty < t < \infty\}$ be a zero mean stationary process with $E[x^2(t)] = S < \infty$ and a power density function $S_x(f)$ such that $S_x(f) > 0 \quad -\infty < f < \infty$ .

In this case it is not possible to consider a proportional-weighted MSE distortion measure since $R(d) = \infty$ for any $0 < d < S$. This could be seen from the lower bound in (16) by representing the process x as a limit (in the mean-square sense) of bandlimited processes with bandwidths $B_n$ and $B_n \to \infty$ as $n \to \infty$ .

However, one can look for a compromise, namely neglecting the signal power density for $|f| \geq B$, for some B, and using the proportional-weighted MSE distortion measure only for the region of signal power density such that $|f| < B$. Then the following mixed distortion measure expressed in terms of the power spectrum of the linear transformation A is suggested:

$$|A(f)|^2 S_x(f) = \begin{cases} \dfrac{S - \delta}{2B} & -B < f < B \\ S_x(f) & |f| \geq B, \end{cases} \qquad (22)$$

where $\delta = 2 \int_B^\infty S_x(f) df$ is an arbitrary small positive number.

Assuming here the existence of $A^{-1}$, the calculating of the rate distortion function of x, under the above weighted MSE distortion criterion, is performed by the calculating of the rate distortion function of y=Ax under the non-weighted MSE criterion.

If x is Gaussian then y is also Gaussian and we use the spectrum of y (the right hand side of (22)) to get

$$R(d) = B \ln \frac{S - \delta}{d - \delta} \ , \qquad (23)$$

for d in the range

$$2BS_x(B) + \delta \leq d \leq S. \qquad (24)$$

In (23), (24) it is assumed that B is large enough, and that $S_x(f)$ is such that $2BS_x(B) + \delta << S$ for $B \to \infty$ .

If x is not Gaussian but $D(x \| \tilde{x}) = D(y \| \tilde{y}) < \infty$, then the following theorem follows directly from [4, Theorem 3].

*Theorem* 3: Under the above conditions and for d that fulfils (24)

$$B \ln \frac{S - \delta}{d - \delta} - D(x \| \tilde{x}) \leq R(d) \leq B \ln \frac{S - \delta}{d - \delta}. \qquad (25)$$

Note that the rate distortion function of time-continuous sources under the MSE distortion measure may behave like $R(d) \approx (1/d)^c, c \geq 1, d \to 0$. Nevertheless, there is no doubt that despite the logarithmic dependency of the rate (25) on d, the rate distortion function under the above mixed distortion measure, in the limited range of d (24), is much higher than that with the non-weighted MSE distortion.

The following example assists in clearing that point.

*Example 2.*

Let us assume that there exists a stationary solution x to the stochastic equation

$$x(t) = x(0) + \int_0^t \alpha [x(s)] ds + \beta w(t), \quad 0 \leq t \leq T \qquad (26)$$

where w is a standard Brownian motion. Since the process that is defined by (26) satisfies $D(x \| \tilde{x}) < \infty$ [4, p523], (25) yields for $d = 2BS_x(B) + \delta$ (section III),

$$R(d) \approx (\frac{\beta}{\pi})^2 \frac{2}{d} \ln \sqrt{\frac{2}{d}} \quad d \to 0. \qquad (27)$$

However, for the non-weighted MSE distortion measure and the same d, R(d) satisfies [4, Example 3]

$$R(d) \approx (\frac{\beta}{\pi})^2 \frac{2}{d} \quad d \to 0. \qquad (28)$$

The rate (28) is smaller than that of (27) for a small d.

## III. APPENDIX

*Proof of theorem 1.*

*Proof of the lower bound.*

For a given MSE distortion vector $\gamma_k^{(n)} = \dfrac{d}{S}\lambda_k^{(n)}$  $k=1,...n$ ,

the following known bound on the information measure between $(X_1^*,...,X_n^*)$ and $(\hat{X}_1^*,...,\hat{X}_n^*)$ is used [4]:

$$I((X_1^*,...,X_n^*),(\hat{X}_1^*,...,\hat{X}_n^*)) \geq$$
$$\sum_{i=1}^{n}\frac{1}{2}\ln\frac{1}{2\pi e\gamma_i^*} + h(X_1^*,...,X_n^*). \quad (A.1)$$

Therefore,

$$R(d) \geq \lim_{n\to\infty}\frac{1}{n}\{\sum_{i=1}^{n}\frac{1}{2}\ln\frac{S}{d} + h(X_1^*,...,X_n^*)$$
$$-\frac{1}{2}\sum_{i=1}^{n}\ln(2\pi e\lambda_i^{(n)})\}. \quad (A.2)$$

The left hand side of (6) follows from (A.2) since

$$D(X\|\tilde{X}) = \lim_{n\to\infty}\frac{1}{n}D((X_1^*,...,X_n^*)\|(\tilde{X}_1^*,...,\tilde{X}_n^*)) =$$
$$\lim_{n\to\infty}\frac{1}{n}\{\sum_{i=1}^{n}\frac{1}{2}\ln\frac{1}{2\pi e\gamma_i^*} + h(X_1^*,...,X_n^*)\}. \quad (A.3)$$

*Proof of the upper bound.*

An upper bound on the rate distortion function so that for every $n$ the distortion vector is given by $\gamma_k^{(n)} = \dfrac{d}{S}\lambda_k^{(n)}$  $k=1,...n$ , has to be shown.

Let $Z = (Z_1,...,Z_n)$ be defined by

$$Z_k = a_k X_k^* + n_k \quad k=1,...,n ,$$

where $a_k = 1 - d/S$ and $n_k, k=1,...,n$ are independent Gaussian random variables, independent of $(X_1^*,...,X_n^*)$ with zero mean and variance $a_k\lambda_k^{(n)}\dfrac{d}{S}$ . Since

$E(Z_k - X_k^*)^2 = \gamma_k^{(n)}$ and (12) holds,

$$R(d) \leq \lim_{n\to\infty}\frac{1}{n}I((Z_1,...,Z_n),(X_1^*,...,X_n^*)) .$$

By first expressing the information in terms of differential entropies and then in terms of divergence we get

$$I((Z_1,...,Z_n),(X_1^*,...,X_n^*)) =$$
$$h(Z_1,...,Z_n) - h((Z_1,...,Z_n)\big|(X_1^*,...,X_n^*)) =$$
$$h(Z_1,...,Z_n) - \frac{1}{2}\sum_{k=1}^{n}\ln 2\pi ea_k\lambda_k^{(n)}\frac{d}{S} =$$
$$\frac{n}{2}\ln\frac{S}{d} - D(Z_1,...,Z_n\|\tilde{Z}_1,...,\tilde{Z}_n).$$

Therefore

$$R(d) \leq \frac{1}{2}\ln\frac{S}{d} - D(Z\|\tilde{Z}) . \quad (A.4)$$

The right hand side of (6) follows from (A.4) since the divergence is non-negative.

*Proof of (7)*

Since Z is a noisy version of $X^*$ , for each d

$$D(Z\|\tilde{Z}) \leq D(X^*\|\tilde{X}^*) . \quad (A.5)$$

Observe that by the definition of Z and (12) the processes $Z, \tilde{Z}$ converge (as $d\to 0$ ) in the mean-square sense to the processes $X^*, \tilde{X}^*$ respectively. Therefore the measures induce by $Z, \tilde{Z}$ (on the real, $\ell^2$ Borel space) converge weakly, as d goes to zero, to the measures of $X^*, \tilde{X}^*$ respectively. Using a known divergence property [10, (2.4.9)]

$$\lim_{d\to 0} D(Z\|\tilde{Z}) \geq D(X^*\|\tilde{X}^*) . \quad (A.6)$$

From (A.4)-(A.6), as $d\to 0$

$$R(d) \leq \frac{1}{2}\ln\frac{S}{d} - D(X\|\tilde{X}) + o(1) \quad (A.7)$$

The asymptotic behavior (7) follows from the left hand side of (6) and (A.7).

*Proof of theorem 2.*

The same sampling technique as described in [1, sec. 4.6.3] to pass from a time-continuous space to a time-discrete space is used. Since this pass preserves both information measure and the proportion of the power density functions $S_X(f), \Phi_X(f)$ , the reconstruction of the process $x = \{x_t, -\infty < t < \infty\}$ with a proportional-weighted MSE per second of d or less is equivalent to the reconstruction of the sequence $X = \{x(kT_s), k=0,\pm1,...\}$ with a proportional-weighted MSE per sample of d or less. It follows that R(d) for $x = \{x_t, -\infty < t < \infty\}$ can be obtained from R(d) for $X = \{x(kT_s), k=0,\pm1,...\}$ by multiplying the rate coordinate of each point on the latter curve by $1/T_s = 2B$ in order to convert from nats per sample to nats per second. Hence, expressions (6), (7) yield (16), (17).

*Proof of example 2.*

To show (27) the following asymptotic behavior of the power spectral density of the stationary process x is used (see [4, (78)] for c=1, r=0)

$$S_x(f) \approx (\frac{\beta}{2\pi})^2\frac{1}{f^2}, \quad f\to\infty . \quad (A.8)$$

Then, by integrating, (A.8) yields

$$\delta \approx 2(\frac{\beta}{2\pi})^2\frac{1}{B} \approx 2BS_x(B), \quad B\to\infty . \quad (A.9)$$

Therefore, for $d = 2BS_x(B) + \delta$ we get

$$d \approx 2\delta, \quad \delta \to 0 . \quad (A.10)$$

From (25), (A.9), (A.10) and $D(x\|\tilde{x}) < \infty$ we get (27).

The asymptotic behavior (28) follows from [4, (27), c=1, r=0].

REFERENCES

[1]  T. Berger, "Rate Distortion Theory – A Mathematical Basis for Data Compression ", Prentice-Hall, Inc. 1971

[2]  R. G. Gallager, "Information Theory and Reliable Communication", Wiley, New York, 1968.

[3]  J. Binia, "On the ε-entropy of certain Gaussian processes", IEEE Trans. Inform. Theory, vol. IT-20, No 4, pp190-196, Mar. 1974.

[4]  J. Binia, M. Zakai and J. Ziv "On the ε-entropy and the rate-distortion function of certain non-Gaussian processes", IEEE Trans. Inform. Theory, vol. IT-20, No 4, pp 517-524, Jul. 1974.

[5]  D. J. Sakrison, "The rate distortion function of Gaussian process with a weighted square error criterion", IEEE Trans. Inform. Theory, vol. IT-14, No 3, pp. 506-508, May 1968.

[6]  J. B. O'Neal, "Bounds on subjective performance measures for source encoding systems", IEEE Trans. Inform. Theory, vol. IT-17, No 3, pp 224-231, May 1971.

[7]  R. A. McDonald and P. M. Schulltheiss, "Information rates of Gaussian signals under criteria constraining the error spectrum", PROC. IEEE, Vol. 52, pp, 415-416, Apr. 1964.

[8]  R. L. Dobrushin and B. S. Tsybakov, "Information transmission with additional noise," IRE Trans. Inform. Theory, Vol. IT-8, pp. 293-304, Sept. 1962.

[9]  A. D. Wyner and J. Ziv, "Bounds on the rate-distortion function for stationary sources with memory", IEEE Trans. Inform. Theory, vol. IT-17, No 5, pp 508-513, Sept. 1971.

[10] M. S. Pinsker, Information and Information Stability of Random Variables and Processes. New York: Holden-Day, 1964.

**Jacob Binia** was born in Jerusalem, Israel, on June 16, 1941. He received his B.Sc., M.Sc., and D.Sc. degrees in electrical engineering from the Technion-Israel Institute of Technology, Haifa, in 1963, 1968, and 1973, respectively.   He joined the Armament Development Authority, Ministry of Defense, Israel, in 1963. From 1973 to 1990 he served first as Head of Signal Processing Group and then as Chief Engineer in the Communications Department, RAFAEL – Electronic Division. From 1992 to 1995 he was part of the National Electronic Warfare Research and Simulation Center.

During the years 1985, 1991, while on Sabbaticals from RAFAEL, he was invited by the Electrical Engineering Faculty at the Technion-Israel Institute of Technology to be a Guest Associate Professor.

He is now with New Elective - Engineering Services Ltd., Israel, where he is employed as a communications systems consultant.